\documentclass[prb,twocolumn,showpacs, amsmath,amssymb, floatfix]{revtex4}

\usepackage{graphicx}
\usepackage{dcolumn}
\usepackage{bm}
\usepackage{subfigure}

\begin{document}

\title{Magnetic properties of the charge density wave compounds $R$Te$_3$, $R$=Y, La, Ce, Pr, Nd, Sm, Gd, Tb, Dy, Ho, Er \& Tm}

\author{N. Ru}
\author{J.-H. Chu}
\author{I. R. Fisher}
\affiliation{Department of Applied Physics, Geballe Laboratory for Advanced Materials, Stanford University, CA 94305 (USA)}

\date{\today}

\begin{abstract}
The antiferromagnetic transition is investigated in the rare-earth ($R$) tritelluride $R$Te$_3$ family of charge density wave (CDW) compounds via specific heat, magnetization and resistivity measurements. Observation of the opening of a superzone gap in the resistivity of DyTe$_3$ indicates that additional nesting of the reconstructed Fermi surface in the CDW state plays an important role in determining the magnetic structure.

\end{abstract}

\maketitle

The $R$Te$_3$ family of compounds ($R$ = Y, La-Sm, Gd-Tm) have generated recent interest as a model charge-density wave (CDW) system.\cite{dimasi_1995, gweon, brouet, brouet_long, laverock, malliakas_2005, malliakas_2006, yao, ru_CDWtransition, johannes} The material is weakly orthorhombic (space group Cmcm) and has a layered crystal structure comprising double layers of nominally square-planar Te sheets separated by a corrugated $R$Te layer.\cite{norling} The relevant parts of the electronic structure are determined by the Te planes, and the Fermi surface (FS) consists of inner and outer diamond-shaped sheets, each doubled due to bilayer splitting, with minimal dispersion in the perpendicular ($b$-axis) direction.\cite{laverock} The Lindhard susceptibility has a distinct peak structure for wave vectors close to 2/7 (5/7) $a^*$ and $c^*$ in the reduced (extended) zone scheme\cite{johannes} (where e.g. $a^*$ = 2$\pi/a$), and indeed all members of the family are found to suffer an incommensurate lattice modulation with wave vector very close to 2/7 $c^*$.\cite{dimasi_1995, malliakas_2005, malliakas_2006, ru_CDWtransition}  Angle Resolved Photoemission Spectroscopy (ARPES) shows that sections of the original FS that are nested by this wave vector are gapped below the critical temperature $T_{CDW}$.\cite{gweon, brouet, brouet_long}  The material remains metallic below $T_{CDW}$\cite{iyeiri, ru_Ce, ru_CDWtransition} and the reconstructed FS can be probed via both ARPES\cite{gweon, brouet, brouet_long} and de Haas-van Alphen measurements.\cite{ru_dHvA} $T_{CDW}$ is found to vary monotonically across the series, with the highest values for members of the family with the largest lattice parameter, this variation being principally ascribed to the effect of chemical pressure on the electronic structure.\cite{ru_CDWtransition} For the four compounds with the smallest lattice parameters ($R$ = Dy, Ho, Er, Tm), remaining portions of the Fermi surface are partially gapped again by a second CDW, which for ErTe$_3$\cite{ru_CDWtransition} and HoTe$_3$\cite{Ho_xray} has been shown to have a wavevector $\approx$ 1/3 $a^*$, transverse to the first. This second transition occurs at a considerably lower temperature than the first, and follows the opposite trend with lattice parameter (i.e. increases with increasing chemical pressure).

Despite the growing interest in this family of compounds, to date the effect of CDW formation on the onset of long range magnetic order (LRMO) has not been considered, principally because the CDW was believed to form at a considerably higher temperature than the N\'{e}el temperature $T_N$. Our recent observation of a second CDW which forms at a considerably lower temperature for the heaviest members of the rare earth series raises the very interesting question of how these two distinct ground states interact (even compete) when $T_{CDW}$ and $T_N$ are on a more level footing. This question motivates an initial survey of the magnetic properties of these materials, and specifically of the magnetic phase diagram, which is the subject of this Brief Report.

\begin{figure}
\includegraphics[width=3.3in]{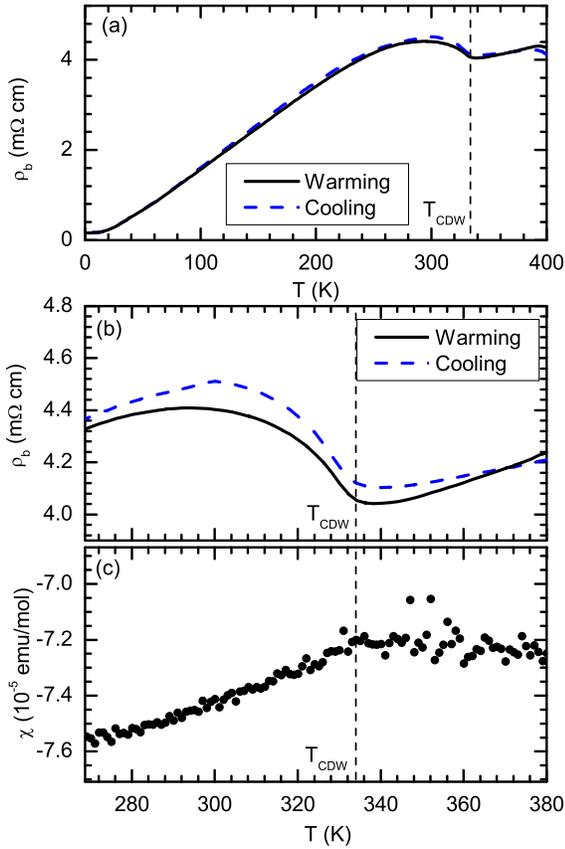}
\caption{\label{fig:YTe3} CDW transition in the non-magnetic compound YTe$_3$. (a) Resistivity for currents oriented along the $b$-axis from 1.8 to 400 K and (b) for the temperature range near $T_{CDW}$. (c) Susceptibility through $T_{CDW}$, measured for $H$ = 5000 Oe oriented in the $ac$-plane.  Vertical lines mark $T_{CDW}$ = 335 K in all three panels. }
\end{figure}
Single crystal samples were grown from a binary melt, as described previously.\cite{ru_Ce} The magnetic susceptibility was measured as function of temperature and field. CDW formation is expected to affect the magnetic susceptibility by decreasing the density of states at the Fermi level, resulting in a reduced Pauli paramagnetism.   In practice, this effect is all but obscured for the magnetic rare earths due to the larger Curie susceptibility.  However, the diamagnetic susceptibility of YTe$_3$ shows a clear downturn on cooling below $T_{CDW}$ (Fig.~\ref{fig:YTe3}).

 As previously reported by Iyeiri \emph{et al.}\cite{iyeiri} for a subset ($R$ = Ce, Pr, Nd, Gd, Dy) of the series, the susceptibilities for the magnetic compounds follow Curie-Weiss behavior and are mildly anisotropic (excepting GdTe$_3$).  Polycrystalline averages of the Weiss temperature are negative for all the compounds, which is indicative of antiferromagnetic exchange interactions.  Magnetic phase transitions were seen for $R$= Ce, Nd, Sm, Gd, Tb, Dy, Ho.  Transitions were not seen for PrTe$_3$, which has a singlet ground state,\cite{iyeiri} or for ErTe$_3$ and TmTe$_3$, for which $T_N$ is below our base temperature of 1.8 K.  A detailed analysis of the susceptibility shows that for most of the series the ground state is reached by a cascade of two or more closely spaced phase transitions, which are more easily seen in the heat capacity.

\begin{figure}
\includegraphics[width=3.3in]{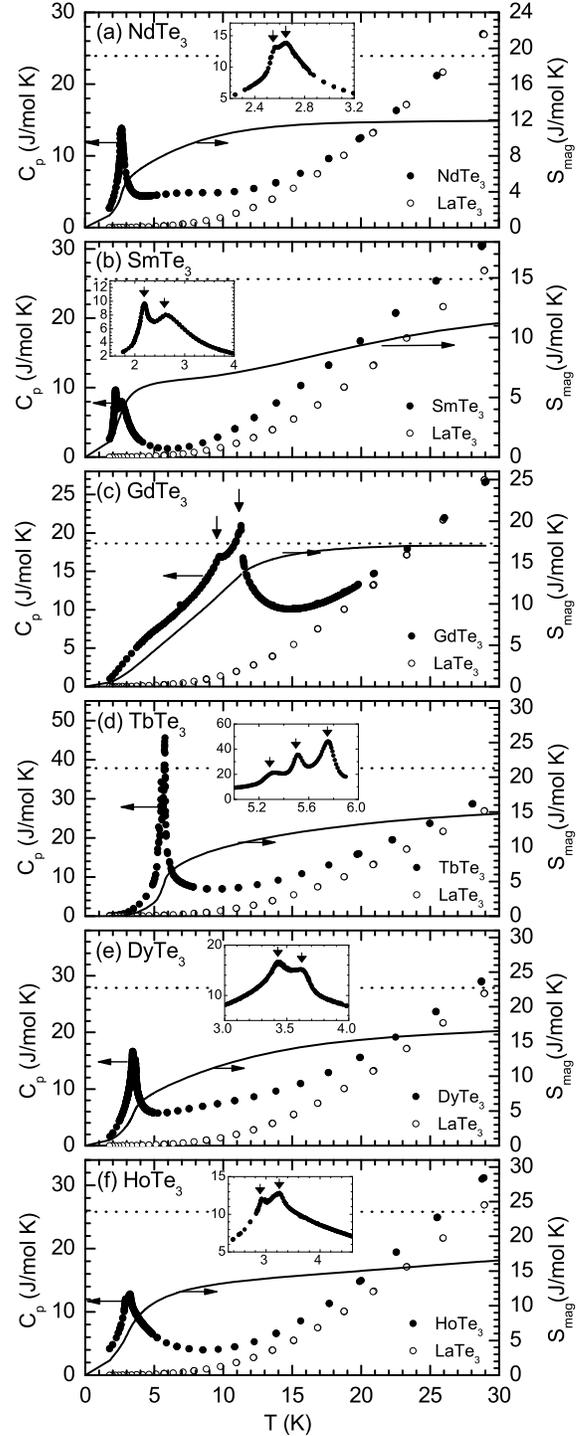}
\caption{\label{fig:heavies_entropy} Specific heat capacity as a function of temperature for (a) NdTe$_3$, (b)SmTe$_3$, (c)GdTe$_3$, (d)TbTe$_3$, (e) DyTe$_3$ and (f)HoTe$_3$ (left axis). Data are shown together with the nonmagnetic analog LaTe$_3$ for comparison. The magnetic entropy $S_{mag}$, estimated as described in the main text, is shown as a solid line (right axis) together with $R$ln(2J+1) (dotted line). Insets show the magnetic phase transitions in more detail, with arrows indicating $T_N$ values.}
\end{figure}

Heat capacity measurements were made using a relaxation technique and are shown in Fig.~\ref{fig:heavies_entropy} for compounds exhibiting LMRO above 1.8 K (Data for CeTe$_3$ appear in Ref.\cite{ru_Ce}).  The heat capacity for the non-magnetic analogue LaTe$_3$ is plotted alongside each data set, to provide an estimate of the phonon contribution to the heat capacity. The onset of LRMO is clearly seen as second-order transitions rising above the phonon background. Each of the compounds exhibit multiple closely-spaced magnetic phase transitions, shown on an expanded scale in the insets to these figures.  For all of the compounds excepting GdTe$_3$ (half filled shell), the heat capacity shows additional contributions from Schottky anomalies associated with CEF splitting of the Hund's rule J multiplets.

The magnetic contribution to the entropy $S_{mag}$, which can be estimated by subtracting the heat capacity of LaTe$_3$, is plotted in each figure as a solid line(right axis).  For reference, $R$ln(2J+1), the entropy for free spins of magnitude J, is shown as a dotted line (right axis). As anticipated, for all of the compounds excepting GdTe$_3$, $S_{mag}$ falls short of this value, indicating the presence of additional CEF splitting on a temperature scale greater than 30 K.  For GdTe$_3$, there is no CEF splitting of the Hund's rule ground state, and as a result  $S_{mag}$ rises to the full value of $R\ln(2J+1)$.

\begin{figure}
\includegraphics[width=3.3in]{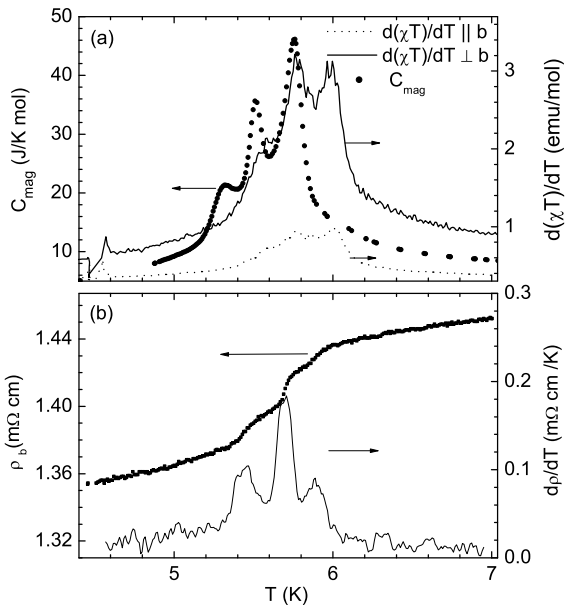}
\caption{\label{fig:Tb_closeup} Successive magnetic phase transitions in TbTe$_3$. (a) Magnetic contribution to the heat capacity (left), and the derivative $d(\chi T)/dT$ of the susceptibility, measured for a field of 1000 Oe oriented parallel and perpendicular to the $b$-axis (right). (b) Resistivity along the $b$-axis (left), with its derivative d$\rho$/dT(right).}
\end{figure}

While most of the compounds in the series exhibit two successive magnetic transitions, TbTe$_3$ is remarkable in having three very closely spaced transitions in an interval of just 0.5 K between 5.3 and 5.8 K. These transitions are also evident in the derivative $d(\chi T)/dT$ of the susceptibility, which is proportional to the heat capacity near a second order phase transition.\cite{FisherME} These derivatives of the susceptibility are shown in Fig.~\ref{fig:Tb_closeup} for two field orientations, together with the same heat capacity data for comparison. Both measurements clearly show three closely spaced phase transitions, with the small difference in peak positions between the heat capacity and susceptibility measurements (0.3 K) ascribed to systematic errors in the thermometry.

The electrical resistivity for TbTe$_3$ is shown in Fig~\ref{fig:Tb_closeup}(b). Clear steps are seen as the temperature is lowered through each of the successive magnetic phase transitions, attributed to the loss of spin-disorder scattering. In this case, the derivative $d\rho/dT$ (right axis), also proportional to the heat capacity for metallic antiferromagnets,\cite{FisherME2} clearly shows the presence of three closely spaced phase transitions. Peak positions are intermediate between those obtained from heat capacity and susceptibility measurements (Fig~\ref{fig:Tb_closeup}(a)), with differences between the three measurements again ascribed to small systematic errors in the thermometry. Excepting DyTe$_3$, resistivity curves for the other compounds in the series either exhibit a similar loss of spin disorder scattering at $T_N$ or show no visible feature.

\begin{figure}
\includegraphics[width=3.3in]{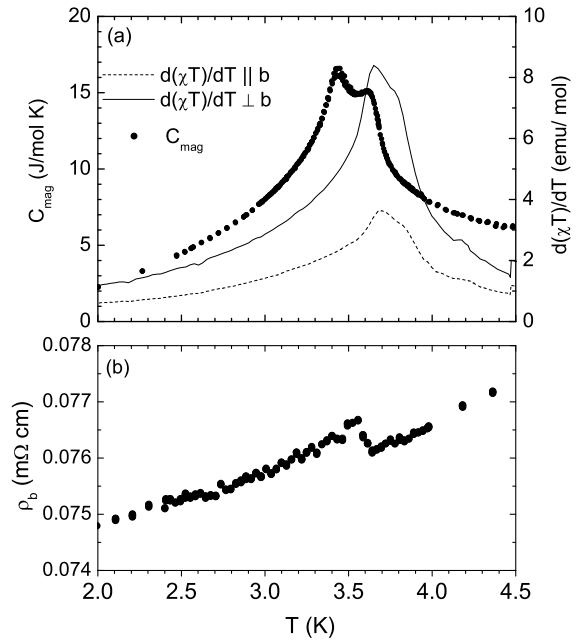}
\caption{\label{fig:dysuperzone} Successive magnetic phase transitions in DyTe$_3$. (a) Magnetic contribution to heat capacity (left), and the derivative $d(\chi T)/dT$of the susceptibility, measured for a field of 1000 Oe oriented parallel and perpendicular to the $b$-axis (right). (b) Resistivity along the $b$-axis, indicating superzone gap formation at $T_N$.}
\end{figure}

 The compound DyTe$_3$ is especially noteworthy since the resistivity shows evidence for superzone gap formation. Two magnetic phase transitions are clearly seen in both the magnetic contribution to the heat capacity and in $d(\chi T)/dT$ (Fig.~\ref{fig:dysuperzone}(a)), with a similar difference between peak positions as seen for TbTe$_3$, due to the small systematic error in the thermometry. In Fig.~\ref{fig:dysuperzone}(b) the $b$-axis resistivity shows a distinct \emph{upturn} on cooling below the upper of the two transitions before dropping due to the loss of spin disorder scattering. Data were collected for many crystals, with only slight differences in the shape of the feature. This behavior is clear evidence for superzone gap formation, implying that the geometry of the reconstructed FS in the CDW state plays a significant role in determining the magnetic ordering wave vector. Since the material has already suffered two CDW transitions, this constitutes a third wavevector gapping sections of the FS. Furthermore, since wavevectors spanning the FS are typically incommensurate with the underlying lattice, this would imply the presence of an incommensurate magnetic structure living in a structure that is \emph{already} modulated by an incommensurate wavevector due to CDW formation. Experiments to directly determine the magnetic structure are underway.

\begin{figure}
\includegraphics[width=3.2in]{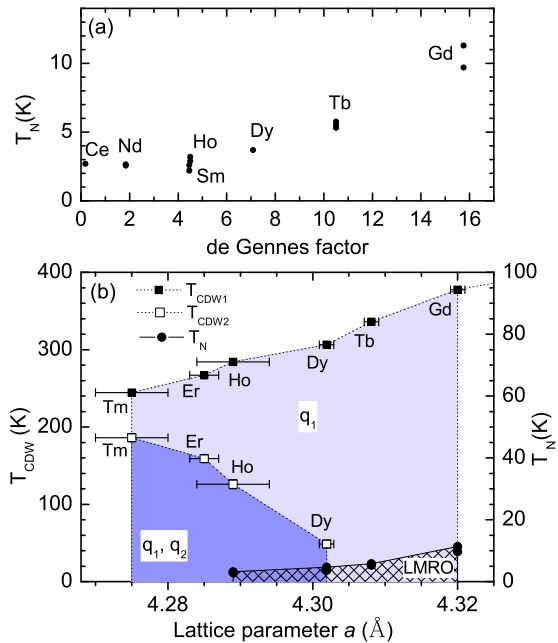}
\caption{\label{fig:degennes}(color online)(a) $T_N$ as a function of the de Gennes factor $(g_J-1)^2J(J+1)$ for all of the magnetic members of the $R$Te$_3$ series that order above 1.8 K. (b) Phase diagram for the heavy rare earth members of the series, illustrating temperature regimes for which the materials exhibit one ($q_1$; light blue) and two ($q_1$,$q_2$; dark blue) CDW wave vectors, and the regimes for which the CDW coexists with long range magnetic order (LRMO; hatched region). }
\end{figure}

N\'{e}el temperatures for the compounds that exhibit a transition above 1.8 K are plotted in Fig.~\ref{fig:degennes}(a) against the de Gennes factor, $(g_J-1)^2J(J+1)$ where $g_J$ is the Land\'{e} g-factor.  Where multiple transitions were found, each transition temperature is plotted separately.  The departure of $T_N$ from a linear dependence on the de Gennes factor can be attributed to the effects of crystal field (CEF) on the Hund's rule multiplet.

In Fig.~\ref{fig:degennes}(b), N\'{e}el temperatures for the heavy members of the series (R = Gd, Tb, Dy, Ho) are plotted as a function of the in-plane lattice constant $a$, together with the CDW transition temperatures $T_{CDW1}$ and $T_{CDW2}$.\cite{ru_CDWtransition, lattice} Of particular relevance, we note that the lower CDW transition $T_{CDW2}$ is rapidly suppressed as $a$ increases. We have previously argued that this effect is directly related to the amount of FS remaining after formation of the first CDW at $T_{CDW1}$.\cite{ru_CDWtransition} Significantly, $T_{CDW2}$ is suppressed somewhere between DyTe$_3$ and TbTe$_3$, raising the distinct possibility that by astute alloying the two order parameters (separately associated with the LRMO and CDW) might be forced to interact on a level playing field. Since the FS plays a key role in establishing both the magnetic ordering in DyTe$_3$ and also the CDW formation - that is, since both effects compete for sections of the FS - it is likely that the phase diagrams for both order parameters will not evolve smoothly between the two end members. Experiments to more closely follow the interaction between these two order parameters for alloys in this range are currently underway.

In summary, in this short paper we have established the outline of the magnetic phase diagram for the rare earth tritelluride family of compounds, in particular highlighting the region for which the onset of long range magnetic order and charge density wave formation occur in a closely spaced range of temperatures. The interplay between LRMO and CDW formation provides another facet to the emerging story of these prototypical CDW compounds.

\section*{Acknowledgements}
This work is supported by the DOE, Office of Basic Energy Sciences, under contract number DE-AC02-76SF00515.

%

\end{document}